# Nanophotonic Modal Dichroism: Mode-Multiplexed Modulators

SUSOBHAN DAS[1], SHIMA FARDAD[1], INKI KIM[2], JUNSUK RHO[2,3], RONGQING HUI[1], ALESSANDRO SALANDRINO[1]*

[1]Department of Electrical Engineering and Computer Science, the University of Kansas, Lawrence KS 66045, USA
[2]Department of Mechanical Engineering, Pohang University of Science and Technology (POSTECH), Korea
[3]Department of Chemical Engineering, Pohang University of Science and Technology (POSTECH), Korea
*Corresponding author: a.salandrino@ku.edu

**As the diffraction limit is approached, device miniaturization to integrate more functionality per area becomes more and more challenging. Here we propose a novel strategy to increase the functionality-per-area by exploiting the modal properties of a waveguide system. With such approach the design of a mode-multiplexed nanophotonic modulator relying on the mode-selective absorption of a patterned Indium-Tin-Oxide is proposed. Full-wave simulations of a device operating at the telecom wavelength of 1550nm show that two modes can be independently modulated, while maintaining performances in line with conventional single-mode ITO modulators reported in the recent literature. The proposed design principles can pave the way to a novel class of mode-multiplexed compact photonic devices able to effectively multiply the functionality-per-area in integrated photonic systems.**

http://dx.doi.org/10.1364/OL.99.099999

Following the success of CMOS electronics, the proliferation of photonic technology relies on the ability of large scale integration to drastically enhance the functionality, lower the per-device cost, and improve the reliability of integrated optical components. As already highlighted in the 2005 International Technology Roadmap for Semiconductors (ITRS) [1], optical interconnects hold promise to meet the ever-increasing requirements of modern telecommunications and processing systems in terms of speed and power consumption. Nanophotonic systems [2] in particular are emerging as one of the most promising technologies for on-chip dense photonic/electronic integration. In fact, scaling down the size of photonic components, compatibly with the constraints imposed by the physics of electromagnetic propagation, can simultaneously increase the bandwidth of operation, decrease the power consumption, and increase functionality-per-area (FPA). During the past decade several photonic-modulator architectures have been developed. Silicon-based structures are attractive from a technological point of view, however, due to the weak electro-optic coefficient of silicon, electro-optic modulators relying on silicon alone [3] require large device footprints on the order of millimeters. The use of high-Q ring resonator structures [4, 5] has been shown to partially circumvent this problem, allowing for the reduction of devices footprint to micron-size dimensions but at the expense of the reduced device bandwidth. Other CMOS compatible architectures featuring materials with superior electro-optic properties have been extensively studied. In particular, graphene-based electro-absorption modulators have been demonstrated [6, 7], and plasmonically-enhanced graphene-based modulators have been proposed [8-10] to improve the performance of such devices. Transparent conductive oxides (TCO) such as indium-tin-oxide (ITO) and aluminum-zinc-oxide (AZO) have recently emerged as promising candidates for both plasmonic [11] and active photonic applications [12, 13]. The wide index-tunability of ITO by doping and by carrier-injection opens the possibility of actively switching between dielectric and plasmonic regimes, thereby enabling the efficient control of optical transmission. At the transition between such regimes – dielectric and plasmonic – lies the epsilon-near-zero condition (ENZ), which could offer significant improvements in the operation of electro-absorption modulators [14].

While materials research has opened new avenues in nanophotonics, many aspects related to electromagnetic design are still to be investigated and fully exploited. Here we consider the problem of increasing FPA from an electromagnetic standpoint and propose novel design principles for nanophotonic modulators that take advantage of modal degrees of freedom in order to effectively double the modulation bandwidth of such devices by introducing additional electromagnetically orthogonal channels that can be modulated independently.

In order to selectively modify the propagation properties – i.e. modal absorption and/or modal index – of different waveguide modes, two different modal characteristics may be used: modal polarization orthogonality, and reduced modal overlap. In order to employ modal polarization for selective modal modulation, a form of tunable polarization dichroism must be introduced in the waveguide core. Let us consider a waveguide described by a permittivity profile $\varepsilon_w(x, y)$ supporting two modes: "mode 1" with fields $\mathbf{E}_1, \mathbf{H}_1$, and "mode 2" with fields $\mathbf{E}_2, \mathbf{H}_2$. Let us also assume that a thin conductive surface with surface conductivity $\bar{\bar{\boldsymbol{\sigma}}}_2$ is inserted in the waveguide cross-section so as to be perpendicular to electric field lines of mode 1 at every point. Under these conditions such conductive surface would only attenuate mode 2, while leaving mode 1 unaffected. Formally one may write the perturbed permittivity distribution of the waveguide as:

$$\varepsilon_0 \bar{\bar{\boldsymbol{\varepsilon}}}(x,y) = \varepsilon_0 \varepsilon_w(x,y) + i\frac{\delta[s_2(x,y)]}{\omega}\bar{\bar{\boldsymbol{\sigma}}}_2 \quad (1)$$

In equation (1) the argument of the Dirac's delta function is the implicit-form equation (i.e. $s_2(x, y) = 0$) of the conductive surface with unit normal $\hat{\mathbf{n}}_1 \parallel \mathbf{E}_1$. The surface conductivity tensor is at every point isotropic with respect to the local tangent plane to the surface $s_2$, and it is null in the perpendicular direction.

Before deriving the modal losses, a word of caution is in order concerning the effects of the conductive surface $s_2$ on the modal field distribution. Such modification of the waveguide's layout in fact changes the electromagnetic boundary conditions, and therefore the modal field distributions $\mathbf{E}_1, \mathbf{H}_1$ and $\mathbf{E}_2, \mathbf{H}_2$ would be altered. Nevertheless if the conductive surface constitutes a small perturbation of the original structure, a perturbation expansion to the first order reveals that only the modal index is affected, while the modal fields are subject only to a second order correction [15]. Under such assumption, the dissipated power per unit length for mode 2 follows from Poynting's theorem, along with the orthogonality relation between modes, and the properties of the tensor $\bar{\bar{\boldsymbol{\sigma}}}_2$:

$$\frac{\partial P}{\partial z} = -\iint \left[\bar{\bar{\boldsymbol{\sigma}}}_2 \cdot \mathbf{E}_2\right] \cdot \mathbf{E}_2^* dx\, dy \quad (2)$$

Equation (2) indicates that mode 2 only experiences attenuation. A similar approach clearly could be adopted to selectively attenuate mode 1, by introducing a thin conductive surface perpendicular to the modal field $\mathbf{E}_2$. Tunability of the aforementioned surface conductivity is necessary in order to realize a mode-multiplexed modulator exploiting the polarization-based modal dichroism just described. Graphene, already used in a number of modulator architectures [6-10, 16-19], would offer ideal characteristics for the realization of the polarization-based mode-multiplexed modulators.

Reduced modal overlap can be exploited to generate "modal dichroism" by introducing a spatially localized absorption mechanism in regions in which only one specific mode concentrates most of its energy, while other modes have small (or ideally zero) amplitude. Notice that such conditions in non-magnetic media can be in principle met for purely transverse electric (TE) modes, but not for transverse magnetic (TM) or for hybrid modes, due to the presence of longitudinal components of electric field. Considering again a waveguide described by a permittivity profile $\varepsilon_w(x, y)$, a loss mechanism can be induced by perturbing the permittivity distribution as follows:

$$\varepsilon_0 \varepsilon(x,y) = \varepsilon_0 \varepsilon_w(x,y) + \frac{i}{\omega}\left[a_1 \sigma_1(x,y) + a_2 \sigma_2(x,y)\right] \quad (3)$$

Let us assume that the parameters $a_1$ and $a_2$ in the imaginary part of the perturbed permittivity profile (3) can be externally modified to assume values between 0 and 1. The power attenuation induced on two waveguide modes $\mathbf{E}_1$ and $\mathbf{E}_2$ can be written as:

$$\frac{\partial P_1}{\partial z} = -a_1 \iint \sigma_1(x,y)\mathbf{E}_1 \cdot \mathbf{E}_1^* dx\, dy - a_2 \iint \sigma_2(x,y)\mathbf{E}_1 \cdot \mathbf{E}_1^* dx\, dy \quad (4)$$

$$\frac{\partial P_2}{\partial z} = -a_1 \iint \sigma_1(x,y)\mathbf{E}_2 \cdot \mathbf{E}_2^* dx\, dy - a_2 \iint \sigma_2(x,y)\mathbf{E}_2 \cdot \mathbf{E}_2^* dx\, dy \quad (5)$$

In order to produce effective modal dichroism the conductivity profiles $\sigma_1$ and $\sigma_2$ must be chosen so as to minimize the cross-terms, i.e. the second term on the right-hand side of equation (4) and the first term on the right-hand side of equation (5). Notice that in general such terms, which are responsible for modulation cross-talk, are always greater than zero, as the argument of the integrals is definite positive. Nevertheless a proper selection of the waveguide modes can greatly facilitate the cross-talk minimization process.

As a simple heuristic guideline for the optimization of the conductivity profiles, the waveguide modes should be selected to have similar order and different symmetry with respect to at least one of the symmetry planes of the waveguide cross-section, as for instance in the case of the $E_{11}^y$ and $E_{21}^y$ modes of a rectangular waveguide [20] in which case the zeros of one mode tend to coincide with the maxima of the other and vice-versa. An additional benefit of such modal choice in terms of minimizing cross-talk is that if the conductivity profiles do not alter the symmetry of the waveguide permittivity distribution $\varepsilon_w(x, y)$, the power exchange between modes of different symmetry is zero. This is the strategy that we adopted in the design of the mode-multiplexed nanophotonic modulator discussed in the next section.

Here for the purpose of illustrating the concept of modal dichroism we consider a mode-multiplexed nanophotonic modulator based on a silicon-on-insulator (SOI) platform for electronic-photonic integration. The configuration of the proposed device is schematically shown in figure 1(a). The optical signal is carried by a multimode silicon ridge waveguide. The active components are three thin plates of ITO of thickness 10 nm that can be individually addressed by means of three gold contacts separated from the ITO structures by an insulating 20 nm layer of $SiO_2$. The dimensions of the ridge waveguide are chosen to be 800 nm in width and 200 nm in height for operating at $\lambda = 1.55 \mu m$ wavelength.

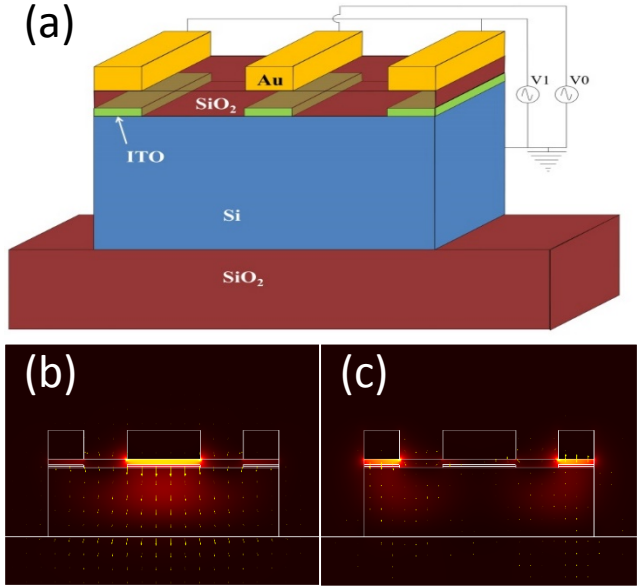

Figure 1. (a) Layout of the proposed mode-multiplexed nanophotonic modulator. (b) Electric field distribution of the $E^y_{11}$ mode for $N_L=N_C=N_0$. (b) Electric field distribution of the $E^y_{21}$ mode for $N_L=N_C=N_0$.

The physical mechanism of modulation in the proposed structure is the voltage-controlled free-carrier absorption in ITO. ITO is a degenerately doped semiconductor with free-carrier concentration that can be tuned by controlling the concentration of oxygen vacancies and interstitial metal dopants. In addition to doping, carrier concentration can also be electrically tuned. Near-unity index changes by carrier injection in ITO have been recently reported [12, 13].

In the layout shown in Figure 1 each of the ITO plates is arranged in a MOS capacitor configuration. With an applied DC potential, the static electric field produces a change in the free-carrier density inside the material by forming an accumulation layer at the ITO-SiO$_2$ interface. The carrier density in the accumulation layer, as obtained from the self-consistent solution of the Poisson's equation for a MOS capacitor is non-uniform, with a profile of the form [21] $n(z) = N_d \exp[\Phi(z)/\Phi_T]$, where $N_d$ is the doping density in the ITO layer, $\Phi_T = k_B T/q$ is the thermal voltage and $\Phi(z)$ is the electric potential at position $z$ within the accumulation layer, with $z=0$ coinciding with the ITO-SiO$_2$ interface. The depth of the accumulation layer is of the order of $d_{acc} \sim \pi L_D/\sqrt{2}$, where $L_D = \sqrt{\varepsilon_0 \varepsilon_{ITO} \Phi_T/(qN_d)}$ is the Debye length in the ITO layer [21].

For our modulator we have considered ITO layers with a doping concentration $N_d = 10^{19} cm^{-3}$. In the top panel of figure 2 we show the calculated carrier density in the accumulation layer for an applied voltage of $V = 16.4V$ across the 20nm SiO$_2$ spacer between the top gold electrodes and the ITO plates. For this voltage value the electric field in the SiO$_2$ spacer is 8.13MV/cm, which is below the breakdown value of ~10MV/cm for this material. The corresponding real and imaginary parts of the ITO permittivity at $\lambda = 1550nm$ [11] in the accumulation layer are shown in the bottom panel of figure 2. At said wavelength and doping concentration, based on the Drude model parameters from Naik et al. [11], the ITO permittivity within the 2.5nm thickness of the accumulation layer varies from the background value of $3.74 + i\,0.00095$ to a value of $-4.59 + i\,0.14$ at the ITO-SiO$_2$ interface.

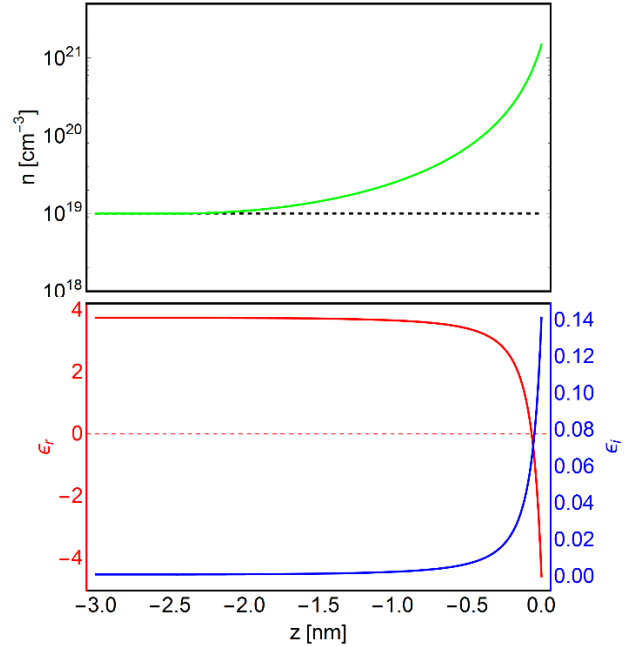

Figure 2. Carrier density (top), and real (red) and imaginary (blue) part of permittivity at λ=1550nm in the accumulation layer of an ITO layer with doping concentration $N_d=10^{19}$cm$^{-3}$ under an applied voltage of 16.4V.

The permittivity profile shown in figure 2 has been used in the numerical modeling of the electromagnetic properties of the proposed modulator. The dimensions of the ITO plates have been optimized through parametric full-wave simulations [22] to maximize the modulation depth while maintaining a low cross-talk between modes, yielding a width of 200nm for the central ITO bar, and a width of 100nm for the lateral bars.

In figure 3 the effectiveness of the proposed modal dichroism mechanism is assessed by studying the modulator performance over the optical communication wavelength range from $\lambda = 1.5\mu m$ to $\lambda = 1.6\mu m$. The Drude model [11] along with the previously discussed MOS model for the carrier density profiles has been used to account for the frequency dependence of the permittivity of the ITO bars. Figure 3(a) shows the propagation losses experienced by the $E^y_{11}$ under the various voltage bias configurations of the ITO bars. The green curve represents the case in which no voltage is applied to any of the ITO bars, and therefore is representative of the device insertion loss. The red curve shows the case in which a voltage $V = 16.4V$ is applied to the central ITO bar, while the lateral bars are left unbiased. In this situation the propagation loss of the $E^y_{11}$ raises to about $0.9\,dB/\mu m$. Finally the blue curve represents the complementary case of a voltage $V = 16.4V$ applied to the lateral bars, while leaving the central bar unbiased: as desired only a marginal increase in the propagation loss of the $E^y_{11}$ mode is observed. The corresponding curves for mode $E_{21}^y$ are presented in figure 3b). As expected, the propagation loss for mode $E_{21}^y$ is strongly affected by the conductive state of the lateral ITO bars, while it remains nearly unchanged by a bias applied to the central ITO bar. These simulations clearly confirm that the proposed structure displays modal dichroism and can be employed as a mode-multiplexed nanophotonic modulator. The proposed modulator shows performances in line with conventional single-mode ITO modulators recently reported in the literature [14], while increasing the functionality per-area by offering the possibility independently modulating two modes.

It is worth mentioning that the theoretical MOS model presented here and in e.g. [14] for the induced carrier density in the ITO plates appears to actually *underestimate* the performance of actual devices[13]. In fact experimental measurements on very similar structures [12, 13] have shown (especially in [13]) performances consistent with much higher carrier densities than the MOS model would predict for the corresponding applied voltages. While some justifications have been proposed [23], more sophisticated models of the accumulation layers in transparent conductive oxides should be explored. This is beyond the scope of this article, since the main focus of this work is to illustrate nanophotonic modal dichroism.

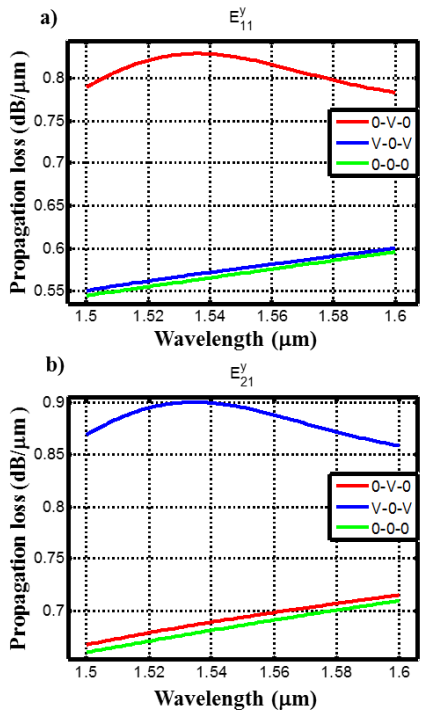

Figure 3. Propagation loss of the (a) $E^y_{11}$ and (b) $E^y_{21}$ modes under various electric bias configurations as indicated in the plot legends.

In conclusion we have introduced the theory and explained the physical principles of operation of modal dichroism in optical and electromagnetic waveguides. We have shown that by exploiting modal degrees of freedom it is possible to increase the functionality-per-area in integrated photonics. Such design principles have been applied to the design of a mode-multiplexed nanophotonic modulator. The proposed SOI device relies on the tunable modal dichroism provided by a patterned ITO film in a triple MOS capacitor arrangement. Different devices may be conceived: while in this letter we concentrated on a nanophotonic modulator, the same structure could serve as mode converter by applying an asymmetric bias among the three ITO bars. The proposed design principles based on either polarization orthogonality or reduced modal overlap can be easily extended to other waveguide configurations, frequencies of operation, physical mechanisms of modulation, and last but not least number of modes. The possibility of selective modulation of two (or more) orthogonal modes within the same structure effectively doubles (or possibly multiplies) the functionality per area that can be achieved. The proposed design principles can pave the way to the realization of novel and densely integrated photonic and optoelectronic architectures.

**Acknowledgements**
A.Salandrino acknowledges AFOSR support through the 2016 Young Investigator Program award FA9550-16-1-0152 and US Army Research Office support through grant W911-NF- 1510377. J. Rho acknowledges the financial support from the National Research Foundation of Korea (NRF-2015R1C1A1A02036464, NRF-2015R1A5A1037668, CAMM-2014M3A6B3063708) funded by the Korean government (MSIP).

## References with titles